\input harvmac
\input epsf.tex
%
% crazy def from hmw
%

\def\IZ{\relax\ifmmode\mathchoice
{\hbox{\cmss Z\kern-.4em Z}}{\hbox{\cmss Z\kern-.4em Z}}
{\lower.9pt\hbox{\cmsss Z\kern-.4em Z}} {\lower1.2pt\hbox{\cmsss
Z\kern-.4em Z}}\else{\cmss Z\kern-.4em Z}\fi}
\font\cmss=cmss10 \font\cmsss=cmss10 at 7pt

% caption def stuff
%Figure Stuff

\let\includefigures=\iftrue
\let\useblackboard=\iftrue
\newfam\black

%Figure Stuff
\includefigures
\message{If you do not have epsf.tex (to include figures),}
\message{change the option at the top of the tex file.}
\input epsf
\def\figin{\epsfcheck\figin}\def\figins{\epsfcheck\figins}
\def\epsfcheck{\ifx\epsfbox\UnDeFiNeD
\message{(NO epsf.tex, FIGURES WILL BE IGNORED)}
\gdef\figin##1{\vskip2in}\gdef\figins##1{\hskip.5in}% blank space instead
\else\message{(FIGURES WILL BE INCLUDED)}%
\gdef\figin##1{##1}\gdef\figins##1{##1}\fi}
\def\DefWarn#1{}
\def\figinsert{\goodbreak\midinsert}
\def\ifig#1#2#3{\DefWarn#1\xdef#1{fig.~\the\figno}
\writedef{#1\leftbracket fig.\noexpand~\the\figno}%
\figinsert\figin{\centerline{#3}}\medskip\centerline{\vbox{
\baselineskip12pt\advance\hsize by -1truein
\noindent\footnotefont{\bf Fig.~\the\figno:} #2}}
\bigskip\endinsert\global\advance\figno by1}
%%%
\else
\def\ifig#1#2#3{\xdef#1{fig.~\the\figno}
\writedef{#1\leftbracket fig.\noexpand~\the\figno}%
%\figinsert\figin{\centerline{#3}}\medskip
%\centerline{\vbox{\baselineskip12pt
%\advance\hsize by -1truein\noindent
%\footnotefont{\bf Fig.~\the\figno:} #2}}
%\bigskip\endinsert
\global\advance\figno by1}
\fi
%
%
% end of caption def

%
% Brian's defns
%

%%%%%%%%%%%%%%%%%%% i stole this ! \def\O{{\cal O}}
%%%%%%%%%%%%%%%%%%% i stole this ! \def\G{{\cal G}} !

\def\T{{\cal T}}
\def\Tp{{\cal T}^\prime}    % this was brookie

            % this too!

\def\slz{SL(2, {\bf Z})}

\
\def\r{{\bf r}}

%  draw box of size #1pt and line thickness #2pt
\def\drawbox#1#2{\hrule height#2pt 
        \hbox{\vrule width#2pt height#1pt \kern#1pt \vrule width#2pt}
              \hrule height#2pt}
% Young tableaux

\def\Asym#1#2{\vcenter{\vbox{\drawbox{#1}{#2}
              \kern-#2pt       % line up boxes
              \drawbox{#1}{#2}}}}

%
% Brookie's defns
%
\def\t#1{{\tilde{#1}}}
\def\frac#1#2{ {#1 \over #2} }

\def\G{{\Gamma}}
\def\a{{\alpha}}
\def\ep{{\epsilon}}

\def\d{{\partial}}
\def\db{{\bar\partial}}
\def\D{{\bf \nabla}_z}
\def\Db{{\bf\nabla}_{\zb}}
\def\DD{{\bf{\nabla}}}
\def\O{{\Omega}}
\def\zb{{\bar z}}
\def\taub{{\bar \tau}}

\def\bt{{\tilde{b}}}
\def\Vt{{\tilde{V}}}
\def\rhot{{\tilde{\rho}}}
\def\taut{{\tilde{\tau}}}
\def\thetat{{\tilde{\theta}}}
\def\bb#1{{\underline{\bf {#1}}}}
\def\twou#1#2#3{{#1}^{#2 #3}}

\def\Guu#1#2{\G^{#1 #2}}
\def\euu#1#2{e^{#1 #2}}
\def\eul#1#2{e^{#1}_{#2}}
\def\fuu#1#2{f^{#1 #2}}
\def\ful#1#2{f^{#1}_{#2}}

\def\aul#1#2{\a^{#1}_{#2}}
\def\epuu#1#2{\ep^{#1 #2}}
\def\epll#1#2{\ep^{#1}_{#2}}

\def\p{{\prime}}

\def\I{{\cal I}}
\def\J{{\cal J}}
\def\K{{\cal K}}
\def\cG{{\cal G}}

\def\uI{{\underline{{\cal I}}}}
\def\uJ{{\underline{{\cal J}}}}
\def\uK{{\underline{{\cal K}}}}

\def\mbR{{\bf R}}

\def\bA{{\bb A}}
\def\bB{{\bb B}}
\def\bC{{\bb C}}
\def\bD{{\bb D}}
\def\bM{{\bb M}}
\def\bN{{\bb N}}

\def\bI{{\bb I}}

%  REFERENCES -----------------------------------------------
%

%%%%%%%%%%%%%%% torus fibration stuff%%%%%%%%%%%%%%%%%%%%%%%%%%%%%

\lref\hmw{
S.~Hellerman, J.~McGreevy and B.~Williams,
``Geometric constructions of nongeometric string theories,''
arXiv:hep-th/0208174.
%%CITATION = HEP-TH 0208174;%%
}

\lref\kodaira{
K. Kodaira,
Annals of Math. {\bf 77} (1963) 563; Annals of Math. {\bf 78} (1963) 1.
}

\lref\scs{
B.~R.~Greene, A.~D.~Shapere, C.~Vafa and S.~T.~Yau,
``Stringy Cosmic Strings And Noncompact Calabi-Yau Manifolds,''
Nucl.\ Phys.\ B {\bf 337}, 1 (1990).
%%CITATION = NUPHA,B337,1;%%
}

\lref\eforf{
C.~Vafa,
``Evidence for F-Theory,''
Nucl.\ Phys.\ B {\bf 469}, 403 (1996)
[arXiv:hep-th/9602022].
%%CITATION = HEP-TH 9602022;%%
}

\lref\syz{
A.~Strominger, S.~T.~Yau and E.~Zaslow,
``Mirror symmetry is T-duality,''
Nucl.\ Phys.\ B {\bf 479}, 243 (1996)
[arXiv:hep-th/9606040].
%%CITATION = HEP-TH 9606040;%%
}

%%%%%%%%%%%%%%%%%%% SUGRA AND HET STUFF %%%%%%%%%%%%%%%%%%%%%%%%%%%%%%%%%

\lref\giani{F. Giani and M. Pernici,
``N=2 Supergravity In Ten-Dimensions,''
Phys. Rev. D {\bf 30}, 325 (1984).
}

\lref\dh{
A.~Dabholkar and C.~Hull,
``Duality twists, orbifolds, and fluxes,''
JHEP {\bf 0309}, 054 (2003)
[arXiv:hep-th/0210209].
%%CITATION = HEP-TH 0210209;%%

}

\lref\nsvi{
K.~S.~Narain, M.~H.~Sarmadi and C.~Vafa,
``Asymmetric Orbifolds,''
Nucl.\ Phys.\ B {\bf 288}, 551 (1987).
%%CITATION = NUPHA,B288,551;%%
}

\lref\nsvii{
K.~S.~Narain, M.~H.~Sarmadi and C.~Vafa,
``Asymmetric Orbifolds: Path Integral And Operator Formulations,''
Nucl.\ Phys.\ B {\bf 356}, 163 (1991).
%%CITATION = NUPHA,B356,163;%%
}

\lref\agm{
P.~S.~Aspinwall, B.~R.~Greene and D.~R.~Morrison,
``Calabi-Yau moduli space, mirror manifolds and spacetime topology  change in
string theory,''
Nucl.\ Phys.\ B {\bf 416}, 414 (1994)
[arXiv:hep-th/9309097].
%%CITATION = HEP-TH 9309097;%%
}

\lref\adp{
K.~Aoki, E.~D'Hoker and D.~H.~Phong,
``On the construction of asymmetric orbifold models,''
arXiv:hep-th/0402134.
%%CITATION = HEP-TH 0402134;%%
}

\lref\jbbs{
J.~Polchinski,
``String Theory. Vol. 2: Superstring Theory And Beyond,''
%\href{http://www.slac.stanford.edu/spires/find/hep/www?irn=4634802}
}

\lref\giveon{
A.~Giveon, M.~Porrati and E.~Rabinovici,
``Target space duality in string theory,''
Phys.\ Rept.\  {\bf 244}, 77 (1994)
[arXiv:hep-th/9401139].
%%CITATION = HEP-TH 9401139;%%
}

\lref\buscher{
T.~H.~Buscher,
``A Symmetry Of The String Background Field Equations,''
Phys.\ Lett.\ B {\bf 194}, 59 (1987).
%%CITATION = PHLTA,B194,59;%%
}

%%%%%%%%%%%%%%%%%%%  flux stuff %%%%%%%%%%%%%%%%%%%%%%%%%%%%%%%%%

\lref\kstt{
S.~Kachru, M.~B.~Schulz, P.~K.~Tripathy and S.~P.~Trivedi,
``New supersymmetric string compactifications,''
JHEP {\bf 0303}, 061 (2003)
[arXiv:hep-th/0211182].
%%CITATION = HEP-TH 0211182;%%
}

\lref\beckerdas{
K.~Becker and K.~Dasgupta,
``Heterotic strings with torsion,''
JHEP {\bf 0211}, 006 (2002)
[arXiv:hep-th/0209077].
%%CITATION = HEP-TH 0209077;%%
}

\lref\ks{
I.~R.~Klebanov and M.~J.~Strassler,
``Supergravity and a confining gauge theory: Duality cascades and
chiSB-resolution of naked singularities,''
JHEP {\bf 0008}, 052 (2000)
[arXiv:hep-th/0007191].
%%CITATION = HEP-TH 0007191;%%
}

\lref\kklt{
S.~Kachru, R.~Kallosh, A.~Linde and S.~P.~Trivedi,
``De Sitter vacua in string theory,''
Phys.\ Rev.\ D {\bf 68}, 046005 (2003)
[arXiv:hep-th/0301240].
%%CITATION = HEP-TH 0301240;%%
}

\lref\gkp{
S.~B.~Giddings, S.~Kachru and J.~Polchinski,
``Hierarchies from fluxes in string compactifications,''
Phys.\ Rev.\ D {\bf 66}, 106006 (2002)
[arXiv:hep-th/0105097].
%%CITATION = HEP-TH 0105097;%%
}

%%%%%%%%%%%%%%%%%%%%%%%%%%%%%%%%%%%%%%%%%%%%%%%%%%%%%%%%%%%%%%%%%%%%%%

\lref\morrison{
D. Morrison,
``Half K3 surfaces,''
Talk given at STRINGS 2002}

\lref\vw{
C.~Vafa and E.~Witten,
``Dual string pairs with N = 1 and N = 2 supersymmetry in four  
dimensions,''
Nucl.\ Phys.\ Proc.\ Suppl.\  {\bf 46}, 225 (1996)
[arXiv:hep-th/9507050].
%%CITATION = HEP-TH 9507050;%%
}

\lref\simeon{
S.~Hellerman, Princeton Journal Club Talk, December 2003
}

\lref\umani{
A.~Kumar and C.~Vafa,
``U-manifolds,''
Phys.\ Lett.\ B {\bf 396}, 85 (1997)
[arXiv:hep-th/9611007].
%%CITATION = HEP-TH 9611007;%%
}

% -----------------------------------------------------------

%
%  Title Page
%

\Title{\vbox{\baselineskip12pt\hbox{hep-th/0404217}
\hbox{UCSD-PTH-04-06}
}} 
{\vbox{\centerline{Constructing Nongeometric Vacua In String Theory}}}
\centerline{Alex Flournoy${}^1$, Brian Wecht${}^2$, Brook Williams${}^3$}
\bigskip
% alex address
\centerline{${}^1$Department of Physics}  
\centerline{Technion, Israel Institute of Technology}
\centerline{Haifa 32000, Israel}
\centerline{\tt flournoy@physics.technion.ac.il}
\vskip.2cm
% brian address
\centerline{${}^2$Department of Physics} 
\centerline{University of California, San Diego} 
\centerline{La Jolla, CA 92093-0354}
\centerline{\tt bwecht@physics.ucsd.edu}
\vskip.2cm
% brook address
\centerline{${}^3$Department of Physics} 
\centerline{University of California, Santa Barbara} 
\centerline{Santa Barbara, CA 93106-9530}
\centerline{\tt brook@physics.ucsb.edu}

\bigskip
\noindent

In this paper we investigate compactifications of
the type II and heterotic string on four-dimensional spaces with 
nongeometric monodromies. We explicitly construct backgrounds which 
contain the ``Duality Twists'' discussed by Dabholkar and Hull \dh.
Similar constructions of nongeometric backgrounds have been discussed
for type II strings by Hellerman, McGreevy, and Williams \hmw.     
We find that imposing such monodromies projects out many
moduli from the resulting vacua and argue that these backgrounds are
the spacetime realizations of interpolating asymmetric 
orbifolds.

%\draftmode
\Date{April 2004}

%
%  PAPER !!!!!!!!!!!!!!!!!!!!!!!!!!!!!!!!!!!!!!!!!!!!!!
%
\newsec{String theory and Geometry}

String theory, in spite of its aspirations as a fundamental theory of
quantum gravity, is for the most part heavily reliant on classical
notions of geometry.  This being said, it is certainly true that
strings and point particles probe classical geometries in
dramatically different ways.
It is well-known that strings can resolve many of the
singularities that plague classical and quantum gravity.  
T-duality establishes a remarkable equivalence between strings
compactified on large tori with those compactified on small tori. 
In more general compactifications, one may take this a step further
and relate geometries of different topologies \'a la mirror symmetry. 
Each of these ideas, however, still has as its foundation classical
geometry.  T-duality and mirror symmetry, though they relate very
different backgrounds, still serve as relations between two classical
geometries, and the stringy resolutions of singularities are
understood in the context of strings propagating in classical
backgrounds.  

There is at least one well-established background which is
intrinsically different from ordinary geometry: One may
consider ``asymmetric orbifolds,'' in which the left-moving modes of
the string see a different geometry than do the right-moving modes.  
In such scenarios, one is no longer able to speak of geometry in a
meaningful way; this is one example of a nongeometric compactification.

One would like to develop a more general framework in which to
discuss string theory backgrounds.  This framework should contain both 
geometric and nongeometric compactifications, and be
intimately related to how strings (rather than point particles) probe
their background.  Such a construction may in fact be a necessary step 
towards a quantum theory of gravity, since one will no doubt have to replace
classical notions of geometry with a quantum alternative.  
As is often the case in physics, symmetries provide us with an
important clue of how to proceed: To construct such 
nongeometric theories, one may take a hint from 
stringy symmetries, such as T-duality, which do not
exist in ordinary quantum field theory.

T-duality is particularly
interesting since, for generic backgrounds, it mixes the metric and
B-field \buscher.  This mixing is an indication that from a string's perspective the
metric and B-field should not be treated as distinct objects but
rather as single field. Mixing between the metric and B-field
is important for the type
of nongeometric compactifications we will investigate in this work; indeed,
this is one sense in which a compactification may be regarded as
intrinsically nongeometric.
Combining the metric and B-field into a single field is certainly not
a new idea; for example, the complexification of the K\"ahler form arises
naturally in string theory.  This complexification allows for mirror symmetry
and leads to interesting physics such as the flop transition \agm.  
It is often the case, however, that the metric and B-field are still treated
as distinct objects which may be combined into a single field.  As will
be emphasized throughout this paper, 
though this field may always be decomposed into
symmetric and antisymmetric parts, as one moves around
nontrivial paths the components of these parts can become intertwined.
By only considering backgrounds where the metric and B-field are
distinct objects, one misses a very large class of
compactifications.  

In the context of the type II string,
the work \hmw\ of Hellerman, McGreevy and Williams (HMW)
exploits stringy symmetries in order to construct a new class of
nongeometric backgrounds. Related ideas, which geometrize the U-duality group,
have also been
discussed in the context of U-manifolds \umani. The main focus of HMW is $T^2$
fibrations over an $S^2$ where the moduli of the fiber
undergo nontrivial monodromies.
The construction of these
nongeometric spaces exactly parallels constructions of
geometric spaces
which exploit the use of geometric symmetry groups; a prominent example
is the construction of K3 via a torus fibration where the fiber has nontrivial
twists by the (geometric) modular group of the torus \refs{\scs}.  
This construction of K3 has played a central role in F-theory
descriptions of type IIB string theory \eforf. The key difference
between the construction of nongeometric backgrounds in \hmw\ and the analogous
construction of K3 is that HMW require that the fiber moduli undergo
monodromies in both the geometric and nongeometric subgroups of the
full perturbative duality group.
The boundary conditions imposed by HMW force the moduli of the theory to vary over 
spaces where the B-field and metric are treated on equal footing 
and cannot be disentangled.
Asymmetric orbifolds are found to be particular limits of these
more general nongeometric spaces. 

Because the base space in \hmw\ is an $S^2$, it does not contain 
any nontrivial 1-cycles.  In turn, the nontrivial monodromies require the 
existence of singularities in the base. If the
base were nonsingular, any closed loop could be shrunk to zero size and
the monodromies would be forced to be trivial. Above each of these
singularities is a degenerate fiber. It
should be noted that although the fibration picture contains
singularities, the total space is smooth.
Had one started with a base that contained nontrivial 1-cycles, it
would no longer be necessary to have a fibration in which the fibers degenerate.
Indeed, since there would now be non-contractible loops, it would be
possible to have nontrivial monodromies without
inducing singularities on the base; such compactifications are the primary focus of this paper.
In particular, we concentrate on $T^2$ fibrations over a $T^2$
base. As will be discussed, the requirement that the base be a $T^2$
forces the moduli of the fiber to lie at fixed points of the imposed
monodromy.  It follows that many of the moduli in these
compactifications are fixed; 
in particular, it is possible to 
fix both the complex structure modulus $\tau$ and K\"ahler modulus $\rho$ of the fiber torus.

HMW focused exclusively on the type II case, and in this work we extend the construction
of nongeometric theories to the heterotic string. In this case, the modular group
is significantly more complicated than that of the type II theories, due to the presence of 
Wilson lines. This greatly enlarges the number of possible monodromies, making a general
analysis quite difficult. As a prototype example, we choose to impose the monodromy
equivalent to T-duality along the fiber torus.

A natural question to now ask is, having fixed the geometric moduli of the theory, in what
sense is the resulting theory nongeometric? Although the answer may at first sound like a 
matter of semantics, we believe that this question hits the very heart of what it means to
be in a nongeometric background. By imposing boundary conditions (i.e. monodromies)
which are intrinsically nongeometric (that is, they mix the metric and the B-field), we arrive 
at a six-dimensional theory with fewer moduli than the corresponding geometric theory. In
the corresponding geometric theory, there exist massless fields corresponding to fluctuations
of the (say) K\"ahler modulus of the torus fiber. In a nongeometric theory, this massless
field is removed from the spectrum, and it is in this sense that the resulting theory is
nongeometric. In other words, to ask whether or not the ten dimensional theory is geometric
or nongeometric is not the appropriate question. Rather, one should ask whether or not
the effective six dimensional theory could be obtained from a geometric theory with geometric
boundary conditions. 

We re-emphasize this point by stressing that a compactification of string theory should
in fact specify two different things: The compact manifold {\bf and} the boundary conditions
around nontrivial cycles. Such boundary conditions (on the bosonic fields) are usually taken
to be periodic, but this is not required. A nongeometric compactification is 
then a compactification with nongeometric boundary conditions. 

It is worth noting that similar compactifications were discussed from a more abstract
perspective in \dh. There, the authors use the Scherk-Schwarz method to prove that
compactifications which include twists of the U-duality group must lie at fixed points; they
also discuss how such theories should have asymmetric orbifold worldsheet descriptions.
By explicitly constructing such a theory (from the spacetime perspective), our
results shed light on this result. In particular, we find that the supergravity equations
of motion require that the moduli lie at fixed points; this condition is not 
{\it a priori} obvious from the supergravity perspective. In addition, we discuss how
finding an asymmetric orbifold description of such a theory is a nontrivial procedure, and
how tightly modular invariance and level matching constrain the possible descriptions.

The outline of this paper is as follows: In Section 2, we review the work of HMW. We focus
specifically on their six dimensional theories coming from nongeometric compacitifications, 
paying special attention to the asymmetric orbifold descriptions of these theories.
In Section 3, we consider type II string theory on nongeometric spaces
which are $T^2$ fibrations over a $T^2$ base. As discussed above,
forcing the base to be a $T^2$ ends up requiring that certain moduli be fixed. 
In Section 4, we consider a similar background for the heterotic string. The situation is
more complicated than the analogous type II theory due to the presence of Wilson lines.
However, as in the type II case, one finds that nontrivial monodromies must have fixed
points. We pick several examples of such monodromies, 
and solve for the allowed backgrounds. 
In Section 5, we discuss interpolating asymmetric orbifolds, and conjecture that these may
correspond to the theories found in Sections 3 and 4.
Finally, in Section 6, we make some concluding remarks and suggest possibilities for future 
work.

\newsec{HMW Review:  Geometric Constructions of Nongeometric String Theories}
\subsec{$T^2$ fibers over an $S^2$ base}
The authors of \hmw\ construct backgrounds by patching
together spaces which are locally fibrations of $T^2$ over $\mbR^2$, 
such that the total space is a nongeometric $T^2$ fibration over
a spherical base.  
We now review this construction in detail.

Consider the dimensional reduction of type II string theory on a $T^2$.
The perturbative duality group of this theory is
\eqn\ottz{
{\cG_{II}} \equiv O(2,2;\IZ) \sim SL(2,\IZ)_\tau \times SL(2,\IZ)_\rho \ .
}
Here $\tau$ refers to the complex structure of the torus, 
\eqn\modtau{ds^2 = \frac{V}{\tau_2} |\tau d\theta_1 + d\theta_2|^2 \ ; \ 
\tau = \tau_1 + i \tau_2 \ , }
and $\rho$ is the complexified K\"ahler modulus which combines the
B-field with the volume modulus of the torus,
\eqn\modrho{\rho = B_{1 2} + i V.}
$SL(2,\IZ)_\tau$ is the geometric modular group which identifies
the modular parameters defining equivalent tori, 
while $SL(2,\IZ)_\rho$ is generated by T-dualities (and a rotation) and 
shifts in the B-field. 
As is well-known, one may change the modulus $\tau$ by any $SL(2,\IZ)$ 
transformation and get a torus physically equivalent to the original.
Thus, when building compact spaces which are fibrations of a $T^2$, 
one may construct closed loops in the base such that upon traversing
the loop, $\tau$ undergoes an $SL(2,\IZ)$ transformation. In this case, one
says that $\tau$ has a nontrivial monodromy along the closed loop. A 
similar story is true for $\rho$; that is, the complexified
K\"ahler modulus may have nontrivial monodromy as well. However, in this case, the
$SL(2,\IZ)$ group comes from the stringy symmetry operations of T-duality  
and shifts in the B-field.

Generically, one expects a $T^2$ fibration to contain
degenerate fibers \syz.  
It is well-known that there are only a finite number 
of different possible degenerations, and each of these induces a different 
monodromy on the moduli. These were classified by Kodaira in 
\kodaira, but we do not review the classification here; we will instead quote
the relevant results when needed.

The construction in \hmw, which is closely related to
\refs{\scs,\eforf}, takes a more active perspective.
Consider a $T^2$ fibered over $\mbR^2$.  
By imposing a particular monodromy around some closed loop, 
a singularity is then induced
on the base below a degenerate fiber. Conceptually, it is clear 
that such a singularity exists:
When considering monodromies around trivial loops 
one may always shrink the loop to zero size, which implies
that the monodromy is also trivial.  
Therefore, since $\mbR^2$ has no non-trivial cycles,
requiring solutions to have non-trivial monodromy forces the
existence of singularities.  The supergravity description breaks down near 
these singularities.  However, thanks to Kodaira's classification,
one is able to describe the
physics near the degenerate fibers in terms of strings on A-D-E
orbifolds.  To construct their backgrounds, HMW assume that far away from 
the singularities the space is fiberwise a solution to the supergravity equations 
of motion, and that these regions
(where supergravity is valid) may be patched together smoothly with
the appropriate A-D-E orbifolds located at degenerate fibers.
Though reasonable, one may question the consistency of such a
construction. There is a nontrivial check which one may perform by
counting the number of hyper, vector, and tensor multiplets and verifying
that they obey $n_H - n_V + 29n_T = 273$, which is necessary for the
theory to be anomaly free. This is indeed satisfied by the nongeometric
construction of HMW.

The appendices of \hmw\ discuss the constraints coming from Type II
supergravity 
in great detail.  In appendix B we rederive (for a slightly different
setup) many of these results.
For now, we will simply state some of the results which come from requiring
a supersymmetric vacuum, i.e. from the condition 
that all SUSY variations of fermions vanish.
The constraint arising from the variation of the dilatino reduces to
\eqn\dilatinoVpre{
{\db B_{12} \over V} = i \chi_6 \bar \del \Phi \ \ ,
}
where $\Phi$ is the dilaton and $\chi_6$ is the six-dimensional chirality
of the conserved spinor.
From the variation of the gravitino along the fiber 
one finds that $\rho$ and $\tau$ must be holomorphic,
$$\db \tau = 0 \quad {\rm and} \quad \db \rho = 0 .$$
Generically such solutions preserve (1,0) supersymmetry in six
dimensions.  For the type IIA theory, solutions with constant $\rho$ preserve (1,1)
supersymmetry and solutions with constant $\tau$ preserve (2,0)
supersymmetry.  This is easy to motivate:  In general, \dilatinoVpre\ fixes
the six dimensional chirality, but for constant $\rho$ the six
dimensional chirality is left unfixed. 
We recall here that a well-known example of such a background arises
from compactifying type IIA on a K3 surface, which does indeed
preserve (1,1) SUSY. 
We can now argue that compactifications with constant $\tau$ preserve
(2,0) SUSY by starting with constant $\rho$ and T-dualizing.

After going to conformal gauge on the base, $ds^2_{base} = e^{2 \phi} |dz|^2$,
the variation of the gravitino along the fiber reduces to
\eqn\gravVfib{
\d \db (\phi - \frac{1}{2} \ln \tau_2 - \frac{1}{2} \ln \rho_2) = 0
\ \ .
}
It easy to show that the modular invariant solution to \gravVfib\ is
\eqn\conffact{
e^{2 \phi} = \tau_2 \Big|\frac{\eta{(\tau)^2}}{\Pi (z - z_i)^{1/12}}\Big|^2 \ 
      \rho_2  \Big|\frac{\eta{(\rho)^2}}{\Pi (z - \tilde{z}_i)^{1/12}}\Big|^2 
\ \ ,}
where $\eta$ is the Dedekind eta function.
The factors of $(z - z_i)$ and $(z - \tilde{z}_i)$ are included so the conformal 
factor does
not vanish at the points $z_i\ (\tilde{z}_i)$ around which there are
$\tau\ (\rho)$ monodromies. 

Far away from any of the degenerations,
$ds^2_{base} \sim |z^{-N/12}dz|^2$, where $N$ is the total number of 
degenerate fibers.  For $N = 24$ the metric is 
$ds^2_{base} \sim |dz/z^2|^2 = |dz^\prime|^2$; $z^\prime \equiv 1/z$.  
It is well known that a two dimensional space which has two coordinate 
patches and a transition function $z^\p = 1/z$ is a sphere. In other words,
the degenerate fibers backreact on the base and cause it to bend, and by
including exactly the right number of singularities, the base curls into
a sphere.  One can show that although the space is compact for $N >
12$, requiring that \dilatinoVpre\ be satisfied at infinity requires
that $N=24$.

It should be noted that the four-dimensional
spaces constructed in this manner are non-singular.  The
``singularities'' on the base are not singularities in the whole space,
but rather an artifact of describing the space as a $T^2$ fibration.

Constructing compact spaces in this manner is not a new idea.
For example, one may describe K3 as a $T^2$ fibration by allowing for
degenerations which cause $\tau$ to undergo monodromies. This
description of K3 has played a central role in
understanding compactifications of F-theory (see e.g. \eforf\ for
discussion).  Since $\tau$ degenerations come from the geometric
symmetries of the torus, the resulting compactification must be purely
geometric.  This is certainly the case for K3.  Similarly, one may
construct compact spaces by considering only $\rho$
degenerations.  Since T-duality exchanges $\rho$ and $\tau$, these
compactifications may also be given a geometric description. 
HMW concentrate on a class of models in which half of
the degenerations induce $\tau$ monodromies and half of the degenerations
induce $\rho$ monodromies.  Since the total order of degenerations
must be 24, the authors of \hmw\ refer these backgrounds as $12 +
12^\prime$ models.  Because the $12+12^\prime$ models have both $\tau$
and $\rho$ degenerations within the same compact space they do not
admit a geometric description.

Let us now try to gain some intuition about $\rho$ degenerations.  Note that
under the monodromy $\rho \rightarrow \rho + 1$ there is not any mixing of the
metric and the B-field.  In turn, these degenerations do not require the 
existence of a new nongeometric background.  Rather, they can be described by
a geometric background in the presence of stringy objects. 
As discussed in \hmw, these degenerations
correspond to NS5-branes.  This is consistent with the fact that
constant $\tau$ solutions in IIA generically preserve (2,0) supersymmetry.
Under T-duality, NS5-branes are 
exchanged with Kaluza-Klein monopoles; the monodromy 
around a KK monopole is $\tau \rightarrow \tau + 1$, as expected.  
This offers an interesting geometrical picture of the 
degenerate fiber:
It is well known that the KK monopole arises from the
dimensional reduction of Taub-NUT space.  Taub-NUT space is often
described as the surface of a ``half cigar."  The degenerate fiber
around which $\tau \rightarrow \tau + 1$ is a ``full cigar''
which has been bent such that the tips of the cigar touch.

In order to better understand the nongeometric aspects of these
compactifications, consider a degenerate fiber around which 
$\rho \rightarrow -1/\rho$.  More explicitly, after 
traversing a closed loop, one has a new metric and B-field with
components
$$B^\prime_{1 2} = \frac{-B_{1 2}}{B_{1 2}^2 + V^2} \ \ , \ \ 
  V^\prime =\frac{V}{B_{1 2}^2 + V^2} \ .$$
It is clear that what was previously thought to be the B-field, $B_{1 2}$,
and what was previously thought to be geometrical, $V$, now mix in a
nontrivial way.
Moreover, the modulus corresponding to rescaling
$\rho$ is fixed.  This is easy to see:
Let $\rho^\p \equiv \lambda \rho$.  As one encircles the degenerate
fiber, $\rho^\p \rightarrow \lambda/\rho \ne -1/\rho^\p$.  The boundary
conditions imposed by the monodromy are no longer satisfied.
One is not free to
rescale the $T^2$; the volume modulus of the fiber is no longer present
in these compactifications. 

One may gain further insight into these spaces by considering an
analogy with classical geometry.  The orbifold $T^4/\IZ_2$ has 16 fixed points
which may be smoothed out by replacing the local geometry
near each fixed point with an Eguchi-Hansen space;
this is the Kummer construction of a K3 manifold.
The radii $\r_i$ of each individual
Eguchi-Hansen space are moduli describing the K3.  The orbifold $T^4/\IZ_2$
is then seen as a particular limit in the K3 moduli space where 
${\bf r}_i \rightarrow 0$ for all $i$.  It is said that the K3 is
the ``blow up'' of the $T^4/\IZ_2$ orbifold.
Analogously, HMW show that there exists a special point in
the moduli space of these nongeometric compactifications which corresponds to an
asymmetric orbifold.  This limit is discussed in detail below.
Since there is a such a point in the moduli space of these
nongeometric spaces which corresponds to asymmetric orbifolds, and
since there is a strong similarity between the construction of these spaces
and K3 manifolds, it is natural to think of these nongeometric
spaces as ``blow ups'' of asymmetric orbifolds.

\subsec{The $12 + 12^\prime$ Orbifold Limit}

Let us now take a closer look at the orbifold point discussed above.
The $12+12^\prime$ orbifold limit is 
\eqn\hmworb{\eqalign{
(T^3\times {\bf R})/ &\IZ_2\times \IZ_2^\prime \quad {\rm with} \quad \cr
\IZ_2 = \I_4 \quad {\rm and} \quad  &\IZ_2^\prime 
= (-1)^{F_L} P_\Delta \I_4 \ .}}
Here $\I_4$ is reflection on all the coordinates of the $T^3$ as well
as the noncompact coordinate $x$, $F_L$ is left-moving spacetime
fermion  number and
$P_\Delta$ is a translation by a distance $\Delta$ along the noncompact 
direction.

\ifig\fixed{
The 8 fixed points at $x = 0$ are resolved $\tau$-degenerations, and
the 8 fixed points at $x = \Delta$ are resolved $\rho$-degenerations.
}{\epsfxsize3.0in\epsfbox{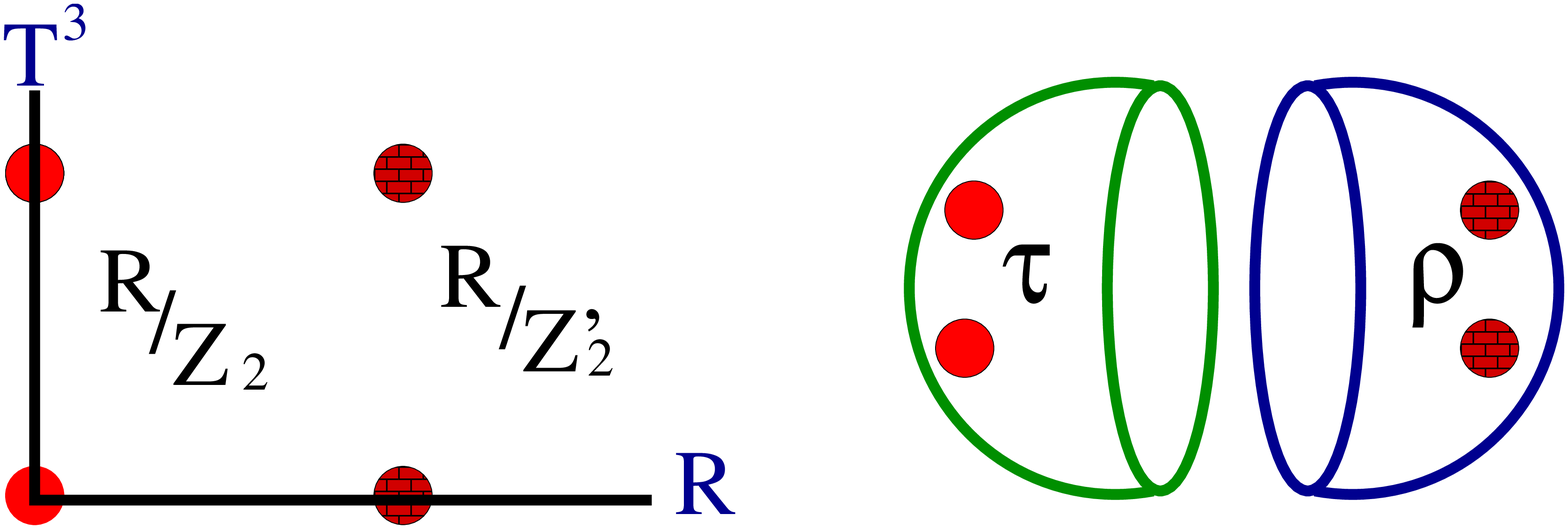}}
\noindent

Associated with the $\IZ_2$ are 8 fixed points located at $x = 0$.
The twisted sector states localized at these fixed points are the same as the 
states in the twisted
sector of IIA on ${\bf R}^4/\I_4$.  In
addition, there are
another 8 fixed points, which come from $\IZ_2^\prime$,
located at $x=\Delta$.  The twisted states on each of these $\IZ_2^\prime$ 
fixed points are
the same as the twisted states of IIA on ${\bf R}^4/\I_4 (-1)^{F_L}$.
As discussed above, K3 has a orbifold limit in which all 16 fixed points are 
described by ${\bf R}^4/\I_4$.  The 8 $\IZ_2$ fixed points in
the $12+12^\prime$ model are identical to those in the orbifold of
Type IIA on a K3. Moreover, since orbifolding type IIA
by $(-1)^{F_L}$ gives type IIB, the
$\IZ_2^\prime$ fixed points are identical to those in the
orbifold limit of Type IIB on a K3. This should not be surprising, since
the $12+12^\prime$ model was constructed by building a compact
space with 12 $\tau$ degenerations and 12 $\rho$ degenerations.  Since
T-duality exchanges $\tau$ degenerations in type IIB with $\rho$
degenerations in type IIA, one can then think of the
$12+12^\prime$ model as gluing together half of a
K3 in type IIA with the T-dual of half of a K3 in type
IIB (see \fixed ).\foot{This should not be confused with the
``half-K3''  spaces in \morrison.}

It turns out that the $12+12^\prime$ orbifold can be reformulated as
an asymmetric orbifold of the K3 orbifold limit.  First note that
\hmworb\ can be rewritten as
\eqn\hmworbtwo{\eqalign{
\big( (T^3\times {\bf R})/ &\IZ_2^{\prime \prime} \big)/ \IZ_2 \quad 
{\rm with} \quad \cr
\IZ_2 = \I_4 \quad {\rm and} \quad  &\IZ_2^{\prime \prime} 
= (-1)^{F_L} P_\Delta \ .}} 
Moreover, 
\eqn\orbequality{\eqalign{
\big( (T^3\times {\bf R})/ \IZ_2^{\prime \prime} \big) &=
\big( (T^3\times S^1_{2\Delta})/ \IZ_2^{\prime \prime \prime} \big) \cr
{\rm where} \ & \ \IZ_2^{\prime \prime \prime} = (-1)^{F_L} s \ .}}
Here $S^1_{2\Delta}$ is a circle of radius $2 \Delta$ and $s$ is a
half shift around the $S^1$. The $T^3$ and the $S^1$ combine to give a
$T^4$, so we find that \hmworb\ can be expressed as
\eqn\orbequalitytwo{\eqalign{
\big( T^4/\IZ_2^{\prime \prime \prime} \big)/ \IZ_2 = 
\big( T^4/\IZ_2\big)/ \IZ_2^{\prime \prime \prime}\ . }} 
Since the $12+12^\prime$ orbifold point can be expressed as a freely
acting asymmetric orbifold of the K3 orbifold point, it is natural to
ask if the $12+12^\prime$ models are simply freely acting asymmetric
orbifolds of the K3.  The authors of this paper are currently
investigating this issue.  
\newsec{Type II Monodrofolds with a $T^2$ Base}

In the remainder of the paper we will discuss new compactifications similar
to those discussed in \hmw.  These compactifications will have a
$T^2$, rather than $S^2$, base.  Since the $T^2$ has non-trivial cycles it is
no longer necessary to have singularities in the base;
this dramatically simplifies the above construction.  
Due to the central role of monodromy in the constructions and in the
backgrounds discussed in \hmw, we have chosen to refer these backgrounds as
``monodrofolds."\foot{Just in case it is not clear, this is a
heterosis of the words monodromy and manifolds,
monodro(my)(mani)folds.} 

In section 2 we briefly discussed supergravity solutions of the
type II string compactified on a $T^2$.  Let us recall some of the
relevant facts:
The supersymmetry transformations of the gravitino $\Psi_{\mu \a}$ and
dilatino $\lambda_\a$ in type II supergravity are 
\refs{\beckerdas, \giani},
\eqn\dilatinovar{
\delta \lambda = ( \G_{[10]} \G^\mu \d_\mu \Phi - 
               \frac 1 6 \G^{\mu \nu \sigma} H_{\mu \nu \sigma} ) \eta = 0
}
\eqn\gravitinovar{
\delta \Psi_\mu = (\d_\mu + \frac 1 4 \O_\mu^{\bM \bN}\Guu \bM \bN )\eta \equiv
\widetilde{\DD}_\mu \eta = 0 \ \ .
}
As usual, we set the expectation values of fermionic fields to zero to obtain a Lorentz invariant
six-dimensional theory, so we need not consider the SUSY variations of the bosonic fields.
These constraints, for a
$T^2$ fibered over an $S^2$, were solved in the appendix of \hmw.  
In the interest of 
completeness, the solutions to \dilatinovar\ and \gravitinovar\ for a
$T^2$ base have been derived in appendix A. 
It turns out that the solutions carry over to the $T^2$ base almost
without alteration:
From the dilatino variation it follows that
\eqn\dilatinoV{
{\db B_{12} \over V} = i \chi_6 \bar \del \Phi \ \ .
}
The gravitino variation along the fiber, $\delta \Psi_I$, reduces to
the holomorphy\foot{
Note that one must replace the complex coordinate $z$ on the sphere
with the complex
coordinate on the torus, $z = \tau \tilde{\theta}_1 +
\tilde{\theta}_2$.} 
 of $\rho$ and $\tau$,
\eqn\sugrafib{ \db \rho = 0 \quad {\rm and} \quad \db \tau = 0 \ , }
and the gravitino variation along the base, $\delta \Psi_i$ becomes
\eqn\sugrabase{
0 = \d \db (\ln \rhot_2 - \ln \rho_2 - \ln \tau_2 ) \ .}
Here $\tau$ and $\rho$ are defined as in \modtau\ and \modrho.
Similarly, we will denote the complex structure and K\"ahler moduli of
the base with $\tilde{\tau}$ and $\tilde{\rho}$.
In general, we will always use tildes to denote parameters describing 
the base.

\subsec{The Metric, B-field, and Ramond-Ramond Moduli}

Unlike the $12 + 12'$ models constructed in \hmw, which had a
singular $S^2$ base, the monodrofolds we are discussing have a
nonsingular $T^2$ base.  This is a tremendous simplification.  
Indeed, as we will show below, this now tells us that $\rho$ and $\tau$ must be constant.  
It follows that the only monodromies allowed are those
with fixed points! The simplest example is
\eqn\monodromy{\eqalign{
\rho \rightarrow \rho \ ,  \ \tau \rightarrow -1/\tau \ \ \ 
(\tilde{\theta}^1-cycle)& \cr
\rho \rightarrow -1/\rho \ , \  \tau \rightarrow \tau \ \ \ 
(\tilde{\theta}^2-cycle)& .
}}
Since $\rho$ and $\tau$ must be constant it follows that 
$\rho = \tau = i$.  All of the fiber moduli are fixed by the monodromy 
\monodromy.  On the surface this compactification looks very much like
$T^2 \times T^2$, where the fiber $T^2$ is square and there is no B-field.
However due to the boundary conditions \monodromy\ there do not exist any 
massless six-dimensional fields coming from moduli of the $T^2$ fiber.

Let us now show why the fields $\rho$ and
$\tau$ must lie at fixed points of the monodromy.  It follows from 
\sugrabase\ that if one allows the moduli of the fiber to vary there
is a nontrivial backreaction on the base.  Such a backreaction is not
possible on a $T^2$.  Let us be more explicit.  
The metric on the base in complex coordinates takes the form  
\eqn\flatmetric{
ds_{base}^2 = 
\frac{\t \rho_2(z,\zb)}{\t {\tau}_2}|\t \tau d\t {\theta}_1+ d\t {\theta}_2|^2 
\equiv \frac{\t \rho_2(z,\zb)}{\t \tau_2} |dz|^2 \ ,}
this is just the usual metric written in terms of the moduli $\t \tau$ and $\t \rho$.
It follows that Ricci scalar $R$ is a total derivative:
\eqn\ricci{\eqalign{
R &= - \nabla^2_{base} \ln \t \rho_2
   = - \frac{\t \tau_2}{\t \rho_2} \d \db \ln \t\rho_2 \cr
  &= - \frac{\t \tau_2}{\t \rho_2} \d \db \ln \rho_2\tau_2 
   = - \nabla^2_{base} \ln \rho_2 \tau_2 \ \ . }}
Going from the first line to the second line of \ricci\ we have used
\sugrabase.
The requirement that the base be a $T^2$ forces 
the Euler characteristic to vanish;
\eqn\chizero{
\chi_{base} = \int_{T^2_{base}} R = 0 \ .}
Since $R$ is a total derivative this is trivially satisfied in
compactifications where $\rho_2$ and $\tau_2$ are single valued.  
However, except in the trivial case where $\rho$ and $\tau$ are 
constants and lie at fixed points of the monodromy, if $\tau$ or
$\rho$ undergoes a nontrivial monodromy the surface term does not vanish.  
We can now see that in order to satisfy \chizero\ and have a nontrivial 
monodromy, $\rho$ and $\tau$ must be constant.

From the supergravity perspective the point is clear:  When performing a
dimensional reduction with a given background, one must make a choice
for the boundary conditions.  In addition to the traditional choice where
the fields are periodic, one may choose boundary conditions in which
the fields undergo a nontrivial monodromy.  The choice of boundary
conditions manifests itself in the field content of the dimensionally
reduced theory, and each different dimensionally reduced theory should
have a consistent world sheet description.  In particular, for the
monodrofolds with a $T^2$ base, we believe that they should be
described by interpolating asymmetric orbifolds \refs{\nsvi,\nsvii}.  These orbifolds will
be discussed in more detail below.  
Indeed, this is the conclusion reached in \dh. There, the authors use
a Scherk-Schwarz argument to prove that stable backgrounds that
include twists in the U-duality group must lie at fixed points of this group,
and should have CFT descriptions as (asymmetric) orbifolds.

As will be discussed in Section 5,
the construction of consistent asymmetric orbifolds is often a laborious
exercise \refs{\nsvi,\nsvii,\adp}.  In contrast, the construction of monodrofolds is substantially
simpler.  The monodrofold construction has the additional advantage that it provides a 
spacetime interpretation of a certain class of asymmetric orbifolds.
Note, however, that this ``spacetime interpretation'' is necessarily 
six dimensional;  since one does not specify the boundary conditions
until one performs a dimensional reduction it does not make sense to
discuss these backgrounds from the higher dimensional perspective.

In addition to $\rho$ and $\tau$, one must also consider the moduli
coming from the Ramond-Ramond sector.
For the type IIA string compactified on a $T^4$, the RR
vector $c_{(1)}$ and the RR three form $c_{(3)}$ each give rise
to four scalars in  six dimensions.
In the monodrofold discussed above, half of the
corresponding moduli are lifted.  To see this one must first ask how
the potentials $c_{(1)}$ and $c_{(3)}$ transform.
An $\slz_\tau$ transformation simply mixes the coordinates of the
torus, and should therefore mix the
potentials coming from branes wrapped on 1-cycles of the fiber
$T^2$.  Such an operation would clearly not affect
branes which wrap both cycles of the torus or branes which do not wrap the
torus at all. We can see that the $\slz_\rho$ action works similarly.
Consider for example the monodromy $\rho \rightarrow -1/\rho$.  
This corresponds to T-dualizing both cycles of the fiber torus
and performing a ninety degree rotation.
The ninety degree rotation simply removes the ninety degree rotation
induced by the two T-dualities.  The T-dualities exchange branes which
do not wrap any cycle of the $T^2$ with branes that wrap the whole
torus.

In particular, the $\slz_\tau$ monodromy, $\T$, acts on on the vectors
\eqn\rrpotsT{
\left (
\matrix{ c_{\theta^1} \cr
         c_{\theta^2} }
\right ) \ {\rm and} \
\left (
\matrix{ c_{\theta^1\thetat^1\thetat^2} \cr
         c_{\theta^2\thetat^1\thetat^2} }
\right ) 
}
via usual matrix multiplication
and the $\slz_\rho$ monodromy, $\Tp$, acts on on the vectors
\eqn\rrpotsTp{
\left (
\matrix{ c_{\theta^1 \theta^2 \tilde{\theta^1}} \cr
         c_{\tilde{\theta}^1} }
\right )  \ {\rm and} \
\left (
\matrix{ c_{\theta^1 \theta^2 \tilde{\theta^2}} \cr
         c_{\tilde{\theta}^2} }
\right )
\ 
}
in the same way.
For definiteness, consider the 2-component object $c_{I}$; this may be either one of the
objects in \rrpotsT. In the
$T^4$ compactification there exist two six-dimensional moduli $\lambda^i$
which correspond to rescalings of the $c_{I}$.  Asking if the
moduli $\lambda^i$ are present in these compactifications is
equivalent to asking if   
\eqn\liftingRR{
\left [ 
\left (
\matrix { \lambda^1 & 0 \cr
          0 & \lambda^2 }
\right ), \T
\right ] = 0 \ .
}
In particular, for the monodromy \monodromy
$$
\T = 
\left (
\matrix { 0 & -1 \cr
          1 &  0 }
\right ) \ .
$$
It follows that $\lambda^1 = \lambda^2$; there is a single modulus
corresponding to an overall rescaling of the 2-vector $c_{I}$.
Similar arguments hold for each of the vectors in \rrpotsT\ and
\rrpotsTp. Thus, this boundary condition projects out half of the moduli
coming from the RR sector.

\newsec{Heterotic Monodrofolds with a $T^2$ base}

The constraints arising from the the dilatino variation \dilatinovar\
and the gravitino variation \gravitinovar\ are the same in the
heterotic string.
There is an additional constraint arising from the
variation of the sixteen gauginos,
\eqn\gauginovar{
\delta \gamma^{\I} = F_{\bM \bN}^{\I} \Guu \bM \bN \eta = 0.
}

Since we are projecting onto spinors of definite four dimensional 
chirality $\chi_4$ on the compact space, \gauginovar\ reduces to the condition that
$F_2$ be (anti)self-dual\foot{As discussed in Appendix A, $\bA$ etc. are tangent space indices along both
the base and the fiber. We also use $I$ as the spacetime index along the fiber,
$i$ as the spacetime index along the base, and $\bI$ as the spacetime index along
both.},
\eqn\wilsonsugra{
0 =  F^{\bA \bB}_{\I} - \frac {\chi_4} 2 \ep^{\bA \bB \bC \bD} F^{\bC \bD}_{\I}
\ \ .}
Converting to spacetime indices and noting that $\d_I A_\bI = 0$ (since
we have assumed that nothing depends on the coordinate along the fiber), this becomes
$F_{IJ}^\I=F_{i j}^{\I} = 0$ and 
\eqn\fiJsugra{
0 =  \d_i A_{J}^{\I} + {\chi_4} \ep_{i j} \ep_{J K}  \d^{j}A^{K}_{\I}.
}
Note that although this is na\"{\i}vely true only for Abelian gauge fields, we can always
choose the Wilson lines to be in a $U(1)^{16}$ subgroup of either $E_8 \times E_8$ or
$SO(32)$; this is due to the fact that the Wilson lines in compact directions
pick up a potential proportional to the square of their commutator when dimensionally reduced.
For generic backgrounds, the only solution to
\fiJsugra\ is $\d_{i} A_J^{\I} = 0$.
Thus, we find that the gaugino variation requires that the gauge fields along the base are
flat and the gauge fields along the fiber are constant.

\subsec{Conventions}

Keeping track of indices becomes even more messy
in the heterotic string than in the type II string.  In an attempt to simplify the equations,
we will work in matrix notation.  As we will describe in detail in the
next subsection, we will discuss the heterotic string in terms of its
embedding into the 26-dimensional bosonic string.
Our main concern will be transformations of the background matrix
$$ E \equiv G + B \ .$$
The rows and columns of $E$ run over the entire compact space (base
and fiber) as well as the internal directions associated with the 
16 left-moving bosons.
We will break the background matrix $E$ into blocks,
\eqn\bkgrnd{
E = \left ( \matrix{  
E_{bb}  &  E_{bf}  &  E_{bI} \cr
E_{fb}  &  E_{ff}  &  E_{fI} \cr
E_{Ib}  &  E_{If}  &  E_{II} 
} \right ) \ \ .}
The subscripts $b$, $f$ and $I$ stand for ``base'', ``fiber'' and
``internal'' respectively.  The subscripts simply label the
directions along which the row and column of a given matrix point.  
They should not be confused with indices: 
they are not raised and lowered with a
metric; repeated labels are not necessarily summed.  
The transpose of a matrix is denoted by a superscript $T$.
When it is necessary to include indices we will label the internal 
directions with indices $\I, \J, \K,\cdots$ and let 
$\uI, \uJ, \uK,\cdots$ 
run over the fiber, base and internal directions.

\subsec{Heterotic T-duality group}
Recall that perturbative duality group for the type II string
compactified on at $T^2$ is
\eqn\ottzagain{
{\cG_{II}} \equiv O(2,2;\IZ) \sim SL(2,\IZ)_\tau \times SL(2,\IZ)_\rho \ .
}
The decompostion of $\cG_{II}$ into $\slz$ subgroups naturally lent
itself to a description in terms of elliptic fibers and the tools
developed in \kodaira.  Such a description was integral to the
construction in \hmw.  In what follows we are interested in the T-duality group
of the heterotic string compactified on a $T^4$ 
(more appropriately, a $T^2$ fibered over a $T^2$);
\eqn\hetgrp{
{\cG_{HET}} \equiv O(20,4;\IZ) \ ,
}
is much more complicated than its type II counterpart $\cG_{II}$.  This structure has 
made the heterotic string extremely interesting from a phenomenological standpoint.
However, for our purposes in this work, the larger duality group significantly complicates 
matters: The duality group is no longer just a product of two $SL(2,\IZ)$ groups. 
In the work that follows we focus on a subgroup of \hetgrp\ which
acts only on the fiber directions.

The heterotic string may be embedded into the 26-dimensional bosonic
string in the following manner (for a review see \giveon):
The bosonic part of the action is simply
\eqn\hetact{
S_{HET} \sim \int d^2 z\, E_{\uI \uK} \d X^\uI \db X^\uK \ ,
}
with the constraint that $X^\I$ be chiral bosons (since the internal 
directions are only along one side of the string); this must be added by hand.
Moreover, the background matrix $E$ must take the following form,
\eqn\hetbkgrnd{
E = \left ( \matrix{  
E_{bb}  &  E_{bf}  &  E_{bI} \cr
E_{fb}  &  E_{ff}  &  E_{fI} \cr
E_{Ib}  &  E_{If}  &  E_{II} 
} \right ) \equiv 
\left ( \matrix{ 
(G + B +  \frac{1}{4}A^2)_{bb} & \frac{1}{4} A^2_{bf}  &  A_{bI} \cr
\frac{1}{4} A^2_{fb}  &  (G + B + \frac{1}{4}A^2)_{ff} &  A_{fI} \cr
0         &                0               &  E_{II} 
} \right ) \ \ .}
The internal background $E_{II}$ is associated with the cartan matrix
$C_{\I \J}$ of the enhanced symmetry group of the heterotic string.
Namely,
$$ E_{\I \J} = 0 \ (\I < \J); \ E_{\I \I} = 1 {\ \rm (NO\ SUM)}; \
E_{\I \J} = C_{\I\J} \ (\I > \J) \ \ .$$
As reviewed in \giveon, this is a consequence of the Narain lattice including the
weight lattice of the gauge group. The indices we have been using are actually
in a dual lattice, which then contains the root lattice of the gauge group; the natural
metric on the root lattice is the Cartan matrix.

One may view this as an compactification on a background where
$$G_{\bI \I}=B_{\bI \I}\ \Longrightarrow\  G_{\I \bI}= - B_{\I \bI} \ \ .$$
It is easy to check that dimensional reduction on the internal space
gives the 10 dimensional heterotic action with Wilson lines
$A^\I_{\bI}$.  For the heterotic string compactified on a $T^2$, the
$\cG_{HET}$ transformations may now be described as $O(20,20;\IZ)$
transformations of the background \hetbkgrnd.   More specifically, the 
$O(20,20;\IZ)$ symmetry group may be generated by integral shifts of
the B-field, 
\eqn\\{B_{ij} \rightarrow B_{ij} + \Theta_{ij} \ \ \ 
(\Theta_{ij} = - \Theta_{ji} \in \IZ) \ ,}
basis changes of the compactification lattice,
\eqn\basis{E \rightarrow M E M^T \ \ \ (M \in O(20,\IZ)) \ ,}
and T-duality.
This is reviewed in great detail in \giveon. As in the type II case we will concentrate on 
examples in which the $T^2$ fiber undergoes these operations.

\subsec{Heterotic Monodrofolds}

We are now in a position to give examples of heterotic monodrofolds with a
$T^2$ base.  As mentioned above we will focus our attention on
monodrofolds in which the fiber alone undergoes monodromies.
Shifts in the B-field and the change of basis have already been made
explicit in the previous subsection. For inversion, one may
follow the steps outlined in \giveon\ to show that 
\eqn\hetmon{\eqalign{
 E_{ff} \rightarrow& E_{ff}^{-1} \cr
 A_{fI} \rightarrow& -E_{ff}^{-1}A_{fI}.}}
This inversion also acts on the other components of $E$ and $A$, but this
action is trivial in the case where $A_{fI} = 0$. 
As it turns out, this will always be the case in our examples.
Since we are only considering transformations along the fiber
 directions we will drop the ``fiber-fiber'' ($ff$) subscripts in what
 follows. Keep in mind that the background $E$ being discussed is a
 $2\times2$ matrix.  This simplifies much of the algebra, and many of
 the formulae which follow are special to the $2\times2$ case.  This
 being said, generalizing to other backgrounds and
 transformations should introduce nothing more than tedium.

In the remainder of this subsection, we will concentrate on examples in
which the above three operations have been performed at most once. Thus, in 
this case there are
really only three different operations we need consider (all other possibilities are equivalent
to special cases of these):
\eqn\jinjur{\eqalign{
1) \qquad &E \rightarrow ME^{-1}M^T + \Theta \cr
2) \qquad &E \rightarrow M(E + \Theta)^{-1}M^T \cr
3) \qquad &E \rightarrow (MEM^T + \Theta)^{-1}. }}
\noindent 

\noindent Note that this is still only a small subset of the
$O(20,20,\IZ)$ monodromies which may be imposed.

We will now go through one example in detail, that of the monodromy 
$E \rightarrow M(E + \Theta)^{-1}M^T$. First note that in this case, the action on
the gauge field along the fiber is $A \rightarrow -(E+\Theta)^{-1}A$, which
is a consequence of composing the three $O(20,20,\IZ)$ transformations in the particular
order we have chosen for this example. Recall that the constraints arising from 
supergravity force the matrix $E_{ff}$ as well as the gauge
fields $A_{fI}$, along the fiber, to lie at fixed points of the
monodromy. Thus, we require
\eqn\hetex{\eqalign{E &= M(E + \Theta)^{-1}M^T \cr
A &= -(E + \Theta)^{-1}A.}}
\noindent It turns out that it is possible to make some general comments, without specifying
the matrices $M$ or $\Theta$. We now run through these arguments in detail.

Taking the trace of the first equation in \hetex, we notice that $\det(E + \Theta)^{-1} = 1$. 
This is easy to see: $\Tr E = \Tr(E+\Theta)^{-1} = \Tr E /
\det(E+\Theta)$.  Thus, either 
$\Tr E = 0$ or $\det(E+\Theta) = 1$. Writing out the matrix $E$, 
\eqn\eff{
E_{ff} = \left ( \matrix{ G_{11} + A^2_{11} & G_{12} + A^2_{12} + B_{12} 
                      \cr G_{12} + A^2_{12} - B_{12} & G_{22} + A^2_{22}} \right ),}
we see that $\Tr E = G_{11} + G_{22} + A^2_{11} + A^2_{22} > 0$, since 
both $G_{11}$ and $G_{22}$ must be positive definite; this is a result of requiring
$\det G > 0$ and our ansatz that the off-diagonal terms in the metric (between the 
fiber and the base) vanish. Thus, $\det(E+\Theta) = 1$, which also implies $\det E = 1$
as a result of our monodromy.

Now, notice that the second equation in \hetex\ implies that, for nonzero $A_{fI}$, the
matrix $(E+\Theta)^{-1}$ has at least one eigenvalue equal to
$-1$. Since basis changes in the orthogonal group \basis\ preserve eigenvalues,
$E = M(E+\Theta)^{-1}M^T$ implies that $(E + \Theta)^{-1}$ and $E$
have the same eigenvalues. Thus $E$ must also
have at least one eigenvalue equal to $-1$.
But since $\det E = 1$, the other eigenvalue must then be $-1$,
implying $\Tr E < 0$. However, since the fiber is purely spacelike this
is not possible.  It follows that the gauge field along the fiber
must vanish; $A_{fI}=0$.

We can now impose some restrictions on the $B$ field along the fiber: Using
\eqn\dete{\det( E + \Theta) = 1 + \Tr E^{-1}\Theta + \det E^{-1}\Theta = 1,}
one can see that $\Tr E^{-1}\Theta + \det \Theta = 0$. Writing
\eqn\thetastuff{
\Theta = \left(    \matrix{
 & -\theta \cr
\theta &  } \right), \qquad \theta \in \IZ}
and noting that
\eqn\einvth{
E^{-1}\Theta = \theta \left ( 
\matrix{ -G_{12}-B_{12} & -G_{22} \cr
G_{11} & G_{12} - B_{12} } \right ), }
we get $-2B_{12}\theta + \theta^2 = 0$, implying
that $B_{12} = \theta/2$. Thus, in the presence of nonzero $\theta$, the moduli
for the $B$-field along the fiber are lifted.

What about moduli for $G$? It turns out that, for generic $M$
with $\det M =1$, all the moduli in the fiber
metric are also fixed by our monodromy. The calculation is straightforward, and yields
\eqn\allmoduli{
G =  {\sqrt{4-\theta^2} \over 2} \left ( \matrix{ 1 & 0 \cr 0 & 1} \right ).}
This implies that $|\theta| < 4$, which means the only possibilities are
$\theta = -1,0,1$. 
For generic $M$ with $\det M = -1$, there are no solutions; such 
transformations are thus not allowed.

Although these results are true in general, there are special cases of $M$ where the above is incorrect.
Taking
\eqn\mone{M = \left ( \matrix{ 0 & 1 \cr -1 & 0} \right ),}
we find that $B_{12} = \theta/2$, and the metric is arbitrary (up to the condition $\det E = 1$).

For 
\eqn\mtwo{M = \left ( \matrix{ 1 & 0 \cr 0 & -1} \right ),}
we find solutions only if $\theta=0$. First,
\eqn\mtwosoli{
G = \left( \matrix{1 & G_{12} \cr
G_{12} & 1} \right), \quad B = \left( \matrix{0 & G_{12} \cr
-G_{12} & 0} \right).}
The second solution for this $M$ is
\eqn\mtwosolii{
G = \left( \matrix{G_{11} & \pm\sqrt{B_{12}^2 + G_{11}^2 -1} \cr
\pm\sqrt{B_{12}^2 + G_{11}^2 -1} & G_{11}} \right), \quad B = \left( \matrix{0 & B_{12} \cr
-B_{12} & 0} \right),}
with $B_{12}$ and $G_{11}$ arbitrary, subject
to the condition that $G_{11}^2 + B_{12}^2 -1 \geq 0$.

Finally, for
\eqn\mthree{M = \left ( \matrix{ 0 & 1 \cr 1 & 0} \right ),}
we find that the only solutions are those with $\theta=0$:

\eqn\mtwosoli{
G = \left( \matrix{G_{11} & 0 \cr
0 &{ 1-B_{12}^2 \over G_{11}}} \right), \quad B = \left( \matrix{0 & B_{12} \cr
-B_{12} & 0} \right)}
with $B_{12}$ and $G_{11}$ arbitrary.

The fixed points of the other two monodromies in \jinjur\ can be figured out in 
a similar fashion. It
turns out that the generic solutions are in fact identical to what we just found:
$A=0$, $B = \theta/2$, and $G$ is as in \allmoduli. This statement is not 
{\it a priori} obvious, and is apparently a consequence of the monodromies we have 
chosen to study. 

\newsec{Worldsheet Constructions}
Throughout this paper we have worked from the spacetime perspective,
exploiting tools from supergravity, in order to obtain new string theory
backgrounds.  If these are indeed legitimate string backgrounds one
should be able to describe these compactifications in terms of a
consistent worldsheet conformal field theory.  It is often very
difficult to find the CFT corresponding to a given string theory
background.  However, we believe that there is a strong indication that 
the monodrofolds with a $T^2$ base discussed in the previous
two sections can be realized as interpolating
orbifolds; similar conclusions are reached by the authors of \dh. To briefly
review, an interpolating orbifold simply combines any
order $n$ discrete transformation $g$ of the worldsheet theory with an
order $n$ geometric shift $s$ along a compact spacetime coordinate to
form a single order $n$ orbifold element $gs$\foot{This is to be
distinguished from $g \times s$.}.  If we start with a theory $H$ on
${\bf R}$ then $H$ on $S^1/gs$ interpolates between the parent theory
$H$ on ${\bf R}$ (as $R \to \infty$) and the pure orbifold $H/g$ on
$S^1_{R/n}$ (as $R \to 0$).  
Note that the monodrofolds discussed above have a similar
interpolating behavior.  Imposing a particular monodromy gives a mass
to various fields in the the theory, which is inversely
proportional to the volume of the base $T^2$.  
In the large volume limit, the fields become effectively
massless and we are left with a standard $T^2$ compactification. As
the volume of the base goes to zero, these fields become infinitely
massive and get projected out of the theory completely.
For a given monodrofold, finding the corresponding interpolating
orbifold boils down to identifying the discrete transformation of the
worldsheet theory corresponding to the particular monodromy of interest.

When the monodromy element corresponds to a nongeometric
transformation of the spacetime
theory, we must consider nongeometric orbifold elements. In many
cases these are transformations acting asymmetrically on the left and
right movers on the worldsheet.  The simplest example of this type is
the monodromy $\rho \to -1/\rho, \tau \to -1/\tau$ which corresponds
to a T-duality transformation on both cycles of the $T^2$ fiber.  
We know that from the worldsheet perspective T-duality is realized as
a reflection on the left-moving fields.  One would na\"{\i}vely
guess that such a monodrofold could be described from the worldsheet
perspective as
\eqn\wrong{
T^2 \times S^1_{2R} / \IZ_2 \ , \ \
\IZ_2 = \I_2^L s \ .}
Here $\I_2^L$ is a reflection on left hand side of the $T^2$ embedding
coordinates, and $s$ is a half-shift of the $S^1$. As will be
discussed below, this is not quite right.

The consistency of interpolating asymmetric orbifolds follows from the 
consistency of both the parent theory $H$ and the pure orbifold $H/g$.  
Unfortunately, the consistency of even pure (i.e. not interpolating)
asymmetric orbifolds is a
highly nontrivial subject.  In the
symmetric case, modular invariance is guaranteed because the theory is
deformed identically on both sides of the worldsheet; this na\"{\i}vely
preserves the level-matching of ground states from the parent
theory\foot{Since the heterotic string is an asymmetric parent theory
a symmetric orbifold may upset level matching.  However, as long as we
embed the spin connection in the gauge connection we are symmetrically
deforming the same set of fields on each side of the worldsheet.}.
For an orbifold that acts differently on the left and right movers,
one generically loses the level-matching of ground states.  Thus, one
must work harder to find asymmetric orbifold actions which lead to
modular invariant theories.  For a given asymmetric transformation
defined in the point group of the worldsheet theory, i.e. an
asymmetric rotation or gauge rotation, one may try to doctor up the
level mismatch by introducing pure translations on the left and right
movers which include geometric and nongeometric coordinate shifts.
Even allowing these modifications does not save most asymmetric
rotations.  For example, there is no way to gauge the rotation of a
single left-moving embedding field (a single T-duality) for any parent
theory and with any combination of shifts.                  

Let's return to the interpolating orbifold described in \wrong. The parent
theory, $T^2/\I_2^L$, was shown not to be modular invariant in \adp.
It was further shown the level mismatch could not be restored by the
inclusion of geometric and nongeometric shifts.  One might be tempted
at this point to question the consistency of these models.  However,
at the level of supergravity we know that these backgrounds are
consistent: recall that all we have done is taken a consistent solution and
imposed consistent boundary conditions.  If there does not exist a
corresponding worldsheet CFT this would be an indication
that the perturbative duality group is broken at the quantum
level.  The authors of this paper do not suspect that this is the case.
A much more conservative conclusion is that
although \wrong\ was the simplest choice, there is another
interpolating orbifold which gives the correct spacetime behavior and
is modular invariant.  Indeed, there are many other interpolating
orbifolds in which the $T^2$ embedding coordinates are antiperiodic
under $2\pi R$ shifts.  The authors of this paper are currently
investigating many of these possibilities.

\newsec{Discussion}

In this paper, we have used nontrivial boundary conditions to arrive at
six dimensional theories with fewer moduli than their counterparts with 
periodic boundary
conditions.  Such nongeometric compactifications seem, at first glance,
to be quite strange.  One might question their relevance in describing 
``real world'' physics.  This being said, there are certainly no
physical or theoretical reasons to suspect the compact dimensions to
have a geometric interpretation.  Moreover, as noted in
\refs{\kstt,\beckerdas} it is often the case that
geometric backgrounds with fluxes are dual to nongeometric
backgrounds.  In fact, it has recently been pointed out that the
Klebanov-Strassler solution \ks, which has played a central role in
the recent discussions of De Sitter vacua in string theory \kklt, 
has a nongeometric dual \simeon. 
To show this, first note that the conifold with NSNS flux may be
decomposed as a $T^3$ fibration over a noncompact three-dimensional base.
The monodromy group of the fiber includes, in addition to geometric
actions, monodromies which shift the periods of the B-field.  T-dualizing
along each of the fibers results in a dual theory in which the
monodromies lie outside of the group generated by the geometric
$SL(3,\IZ)$ transformations and B-field shifts.  The new monodromies
include nongeometric monodromies, i.e. monodromies which change
volumes of the $2$-cycles and the dimensionality of branes
wrapped on various cycles.  

It is becoming increasingly clear that nongeometric backgrounds are an
indeed important ingredient of string theory.  Such vacua, however, have not 
been thoroughly explored, leaving many open questions for future work.
This work is a step in exploring these vacua. 
Of course, we have not here studied the most general 
nongeometric
boundary conditions possible. In particular, our discussion of the 
heterotic
string imposes monodromies which are a very small subset of the 
$O(20,4,\IZ)$ duality
group. We find that in general, these monodromies must have fixed 
points and
that in all cases the Wilson lines along the fiber are forced to 
vanish. A natural question
is to ask whether one can see this (from a spacetime perspective) in 
more general
nongeometric spaces. It would also be interesting to find backgrounds 
in which the Wilson
lines are nonvanishing.

As mentioned in section 5, one would like to have a worldsheet 
description of nongeometric
theories. A natural guess for such a description is that it is an 
interpolating asymmetric orbifold. However,
the construction of asymmetric orbifolds is known to be a very delicate 
procedure
\refs{\nsvi,\nsvii,\adp}. Finding such a description would be a
big step in understanding general nongeometric backgrounds.
Another interesting question would be to find a string dual of the
nongeometric compactifications in \hmw\ or this paper. Since the $12 + 
12^\prime$ is
very similar to an orbifold of K3 (see Section 2.2), it may be possible to do this 
using the techniques of \vw. One would suspect that the dual of the IIA 
theory would be some kind of nongeometric
compactification of the heterotic string, possibly one of the models 
discussed here.

Another thing that one should note is that in this paper we have not checked that these
nongeometric backgrounds are anomaly free.
However, as discussed in Section 5, all we have done is taken
a consistent theory and imposed consistent boundary conditions. 
We have no reason to suspect that, at the level of supergravity, this should
result in an inconsistent theory. 
To better understand the nongeometric theories presented in this paper, it would be interesting
to compute the spectrum and show explicitly that they are anomaly free. 

Finally, we mention that in general, nongeometric compactifications of 
string theory are not
well-understood. It may be possible to find such compactifications 
which are neither
monodrofolds nor asymmetric orbifolds, but something else entirely. 
There are many possibilities
for string vacua among such spaces, and by analogy with the 
nongeometric examples
discussed here, one could hope that these also project out moduli. In 
order to obtain
a more  complete picture of string vacua, it is essential that we study
such nongeometric spaces.

\bigskip
\centerline{\bf{Acknowledgements}}

We are pleased to acknowledge useful discussions with Oren Bergman, Amihay Hanany,
Gary Horowitz, Ken Intriligator, and Joe Polchinski.
We would like to give special thanks to Simeon Hellerman who has been a
tremendous help throughout this project.   For those who are
interested, Nick Halmagyi did have a very nice day.
The work of A.F. was supported by the Israeli Science Foundation under grant number
101/01-1. The work of B. Wecht was supported by Department of Energy grant DOE-FG03-97ER40546.
The work of B. Williams was supported by National Science Foundation grant PHY-0244764.

%\vfill\eject

\appendix{A}{Conventions} 

Unfortunately with the total space, the total compact space, the base space, 
the fiber space and all of the respective tangent spaces, it becomes quite ugly
trying to keep track of all the different kinds of indices.  (The situation 
is even worse for the heterotic string!)  We will try our best to keep things 
clear.  In general, boldfaced 
underlined letters correspond to the entire compact space.  Lowercase
letters go with the base and uppercase letters go with the fiber.  Moreover,
the letters at the beginning of the alphabet (A, B, C,$\cdots$) correspond to the 
to the tangent space indices and letters in the middle 
of the alphabet (I, J, K,$\cdots$) correspond to spacetime indices.  

More explicitly we choose the following conventions for our coordinates:
$X^\mu$ are the coordinates on all of spacetime.
$\tilde{\theta}^i$ are the coordinates on the $T^2$ base,
and $\theta^I$ are the coordinates on the $T^2$ fiber.

Let $\bb{M},\bb{N},\bb{P},\cdots $ be the entire set of tangent space
indices;
let $a,b,c,\cdots $ be the tangent space indices corresponding to 
$\tilde{\theta}^i$;
let $A,B,C, \cdots$ be the tangent space indices corresponding to
$\theta^I$; let $\bb{A},\bb{B},\bb{C},\cdots $ be indices that run over
both $a$ and $A$.  We take
the $\theta^I$ coordinates (and similarly for $\tilde{\theta}^i$) 
to have constant periodicity $\theta^I \sim
\theta^I + n^I, n^I \in \IZ$.

Additionally, tildes denote coordinates and fields parameterizing the base 
torus: $\tilde{\theta}^i, \taut, \rhot, \bt, \Vt$. 
$e_i^a\ (f_I^A)$ is the vielbein on the 
base (fiber). Finally, we use $\widetilde{\DD}$ to denote 
the generalized covariant derivative 
(i.e. the one corresponding to the generalized spin connection) 
acting on spinors and $\DD$ 
to denote the ordinary covariant derivative.

These conventions are consistent with \hmw, though we will make one small change.
In \hmw\ the generalized spin connection was defined to be
$$ \Omega_\mu^{ \bb {MN} } \equiv
{\omega_\mu}^{ \bb {MN} } +
{H_\mu}^{ \bb {MN} } \G_{[10]} \ \ \  \ {\rm (OLD)} \ .
$$
We will use the conventions of \jbbs\ where
\eqn\genspin{\Omega_\mu^{ \bb {MN} } \equiv
{\omega_\mu}^{ \bb {MN} } +
\frac{1}{2}{H_\mu}^{ \bb {MN} } \G_{[10]} \ \ \ \ {\rm (NEW)} \ .
}
This new definition of $\Omega_\mu^{ \bb {MN} }$ effectively redefines
the B-field such that $\rho \equiv b + i V$ is holomorphic 
(rather than $b + i V/2$).  Here we are using
$b \equiv B_{12}$ and
$V\equiv\sqrt{{\rm det} M}$, rather than their respective periods, since it is 
these objects which change under the Buscher rules.  This being said, in the 
compactifications we are considering these are constants, and since the torus 
coordinates have period = 1, $B_{12}$ and $V\equiv\sqrt{{\rm det} M}$
are equal to their periods.

\appendix{B}{Type II Supergravity on a $T^2$}

The supersymmetry transformations of the gravitino $\Psi_{\mu \a}$ and
dilatino $\lambda_\a$ in type II supergravity are 
\refs{\beckerdas, \giani},
\eqn\dilatinovar{
\delta \lambda = ( \G_{[10]} \G^\mu \d_\mu \Phi - 
               \frac 1 6 \G^{\mu \nu \sigma} H_{\mu \nu \sigma} ) \eta = 0
}
\eqn\gravitinovar{
\delta \Psi_\mu = (\d_\mu + \frac 1 4 \O_\mu^{\bM \bN}\Guu \bM \bN )\eta \equiv
\widetilde{\DD}_\mu \eta = 0 
}
The constraints coming from \dilatinovar\ and \gravitinovar\ for a
$T^2$ fibered over a $S^2$ were solved in the appendix of \hmw.  
These constraints should carry over to the $T^2$ base, but for
completeness we will now
rederive some of these for this case.

The derivation of the constraint coming from the dilatino variation is
exactly as for the $S^2$ case in \hmw. Thus, we will simply remind the
reader that, as stated in section II, \dilatinovar\ reduces to
\eqn\dilatinoV{
{\db b \over V} = i \chi_6 \bar \del \Phi \ \ .
}
As shown in \hmw, for a general two dimensional base
$\CB_2$, 
the killing spinor equations along the fiber, $\widetilde{\DD}_I \eta = 0$, imply 
that \foot{Note that this differs from HMW by a factor of 1/2.  This is 
due to \genspin.}
\eqn\sgrafiber{ \euu i a {\d_i V}
                + \chi_6 \twou \ep a b \euu i b \d_i b = 0 \ \ .} 
For a spherical base $\CB_2 = S^2$, we may take $\euu i a \propto
\delta^{i a}$, and these equations reduce to the Cauchy-Riemann
equations, $\db \rho = 0$.  Moreover, it can be shown $\db \tau = 0$.  
This can be
seen directly, as in \hmw, or seen as a consequence of the fact
that T-duality interchanges $\rho$ and $\tau$. 
For $\CB_2 = T^2$, \sgrafiber\ again reduces to holomorphy of
$\rho$.  Note however that holomorphy is now with respect to the complex
coordinates on the torus,
$ z = \tau \theta_1 + \theta_2, \ \zb = \taub \theta_1 + \theta_2 $.
  
In order to solve the killing spinor equations along the base, 
$\widetilde{\DD}_i \eta = 0$, we first note that, for the ansatz and basis of gamma 
matrices used in \hmw,
$$\Guu A B = \epuu A B \sigma^3, 
\ \Guu a b = \epuu a b \sigma^3 \ , 
\ \O_I^{A B}=\O_I^{a b}=\O_i^{a A} =0 .$$
Going to complex coordinates, we see that the equation 
$0 = [ \widetilde{\DD}_z, \widetilde{\DD}_{\zb} ]  \eta$ reduces to
\eqn\commutator{
\eqalign{ 0
&= \d \O_{\zb}^{\bM \bN} \Guu \bM \bN - \db \O_z^{\bM \bN} \Guu \bM \bN \cr
&=  \d [ (\epuu A B \fuu A I \Db \ful B I) +
         (\epuu a b \euu a i \Db \eul b i) +
    (\frac{\db b}{V}) ] \cr
&\ \ \ \ \ \ - \db [ (\epuu A B \fuu A I \D \ful B I) + 
                     (\epuu a b \euu a i \D \eul b i) +
    (\frac{\d b}{V}) ]
}}
Here we have used the fact that the vielbein along the fiber satisfies
$$\epuu A B \epll I J \fuu A I \fuu B J 
= 2 det[\fuu A I] = 2 V^{-1}\ \ .$$
We will now consider each term (minus its complex conjugate)
individually.  Before moving on, though, it will be useful to note that
\eqn\christofzero{ 
\G^K_{I J} = \G^k_{i J} = \G^K_{i j} = 0 \ \ .}  
Moreover, the amount of algebra may be reduced by rescaling 
$\ful A I$ (and similarly for $\eul a i$) since 
$$\epuu A B\fuu A I \Db \ful B I 
   =\epuu A B \a^{A I} \Db \a^A_I \ \ ;$$
here $\a^A_I \equiv h(z,\zb) \ful AI$ for an arbitrary function
$h(z,\zb)$.  We have chosen to work with the explicit
(rescaled) vielbein $\aul A I$:
\eqn\viel{
\eqalign{\aul 1 1 = |\tau|^2 \ \ 
    &\ \ \aul 1 2 = \tau_1 \cr
         \aul 2 1 = 0 \ \
    &\ \ \aul 2 2 = \tau_2
     \ \ .}}

It is easy to show that 
$\epuu A B \fuu A I \Db \ful B I = \epuu A B \fuu A I \db \ful B
I$.
From the holomorphy of $\tau$ it follows that
\eqn\tauterm{
\d (\epuu A B \fuu A I \db \ful B I) - \db (\epuu A B \fuu A I \d \ful
B I) = 2 i \frac {\d \tau \db \taub} {|\tau - \taub|^2} 
= - 2 i \d \db \ln \tau_2 \ .}
Now consider the the second term on the RHS of \commutator.  We are
free to chose a gauge in which $\taut$ is constant.  After doing this, 
$\epuu a b \euu a i \Db \eul b i 
= - \epuu a b \euu a i \G^j_{\zb i} \eul b j$.  A little algebra
yields
\eqn\rhotterm{
\d (\epuu a b \euu a i \Db \eul b i) - \db (\epuu a b \euu a i \D \eul b i) 
= 2i \d \db \ln \rhot_2 \ \ .}
Finally, note that the forth term is greatly simplified by the
holomorphy of $\rho$.
\eqn\rhoterm{
\d (\frac{\db b}{V}) - \db (\frac{\d b}{V}) = 
- 2i \d \db \ln\rho_2} 
Putting all of this together, one finds that
\eqn\sugrabase{
0 = \d \db (\ln \rhot_2 - \ln \rho_2 - \ln \tau_2 ) \ \ .}
As stated, this is the same solution as found in \hmw\ for a
spherical base.

\listrefs
\end